\documentclass[11pt]{article}

\usepackage{amssymb}
\usepackage{amsmath}
\usepackage{graphicx}
\usepackage[table]{xcolor}
\usepackage[T1]{fontenc}
\usepackage{setspace}
\usepackage{array}
\usepackage{tabularx} 
\usepackage{authblk}
\usepackage[margin=1in]{geometry}
\usepackage[authoryear,round,compress]{natbib}
\usepackage{booktabs}
\usepackage{rotating}
\usepackage{booktabs} 
\usepackage{caption}
\usepackage{multirow}
\usepackage{adjustbox}
\usepackage{geometry}
\usepackage{hyperref}
\usepackage{setspace}
\usepackage{float}
\usepackage{placeins}
\usepackage{pdflscape}
\usepackage{caption}
\usepackage{bm}
\usepackage{subcaption}
\usepackage{comment}

\DeclareCaptionFormat{vertical}{\rotatebox{90}{#1#2#3}}

\usepackage{amsmath}
\usepackage{geometry}
\usepackage{hyperref}
\usepackage{setspace}
\usepackage{booktabs}

\author[1,2]{Piersilvio De Bortoli}
\author[2]{Davide Ferrari}
\author[2,3]{Francesco Ravazzolo}
\author[4,5]{Luca Rossini}

\affil[1]{University of Trento}
\affil[2]{Free University of Bozen-Bolzano}
\affil[3]{BI Norwegian Business School}
\affil[4]{University of Milan}
\affil[5]{Fondazione Eni Enrico Mattei}

\begin{document}

\title{Model selection confidence sets for time series models with applications to electricity load data}

\date{}


\maketitle

\doublespacing

\begin{abstract}
This paper studies the Model Selection Confidence Set (MSCS) methodology for univariate time series models involving autoregressive and moving average components, and applies it to study model selection uncertainty in the Italian  electricity load data. Rather than relying on a single model selected by an arbitrary criterion, the MSCS identifies a set of models that are statistically indistinguishable from the true data-generating process at a given confidence level. The size and composition of this set reveal crucial information about model selection uncertainty: noisy data scenarios produce larger sets with many candidate models, while more informative cases narrow the set considerably. To study the importance of each model term, we consider 
numerical statistics measuring the frequency with which each term is included in both the entire MSCS and in Lower Boundary Models (LBM), its most parsimonious specifications.
Applied to Italian hourly electricity load data, the MSCS methodology reveals marked intraday variation in model selection uncertainty and isolates a collection of model specifications that deliver competitive short-term forecasts while highlighting key drivers of electricity load like intraday hourly lags, temperature, calendar effects and solar energy generation.
\end{abstract}

 {\it Keywords:}  model uncertainty, model selection confidence set, lower boundary models, electricity load forecasting, time series.

\section{Introduction}

Model selection is an important tool for practitioners and policymakers as it enhances the interpretability and practical relevance of statistical models. This is particularly evident in time-series analysis, where competing models can deliver similar in-sample fits while producing substantially different forecasts \citep{stock,Rossi}. More often than not, however, the data do not clearly support a unique model, so model selection uncertainty becomes a central concern.

This issue is important in commodity markets, particularly for partially storable commodities such as electricity, where country-level load modeling has gained prominence. Electricity load exhibits pronounced intra-day variation and responds to a broad set of drivers, including weather conditions \citep{station,recency,lowvoltage,ireland}, economic indicators \citep{FEZZI,LiddleHasanov2024Correction}, and seasonal patterns \citep{taylor2,DORDONNAT2008566,seasonal,bookseason}, making the identification of a single best specification difficult \citep{hong2019gefcom2017,dong2025stelfsurvey}.
 Moreover, the optimal model structure may change over time as demand patterns evolve (concept drift) and regulatory reforms  reshape consumption behavior \citep{Enrich2024TOUSpain,azeem2024tla,cao2025opdf}. Understanding and controlling this variability matters because even modest forecasting improvements can reduce operating costs by improving unit commitment and dispatch, informing reserve requirements under net-load uncertainty, and lowering congestion-management needs \citep{pierro2020residualload,wang2023valueprob,bernecker2025congestion,hasan2025lfreview}. This becomes even more prominent during extreme load conditions (e.g., heat- or cold-driven demand spikes), which are important to predict, and model comparisons can depend on how performance is evaluated when attention is placed on extremes \citep{lerch2017forecasters}.

\begin{figure}[h!]
    \centering
    \includegraphics[width=0.5\linewidth]{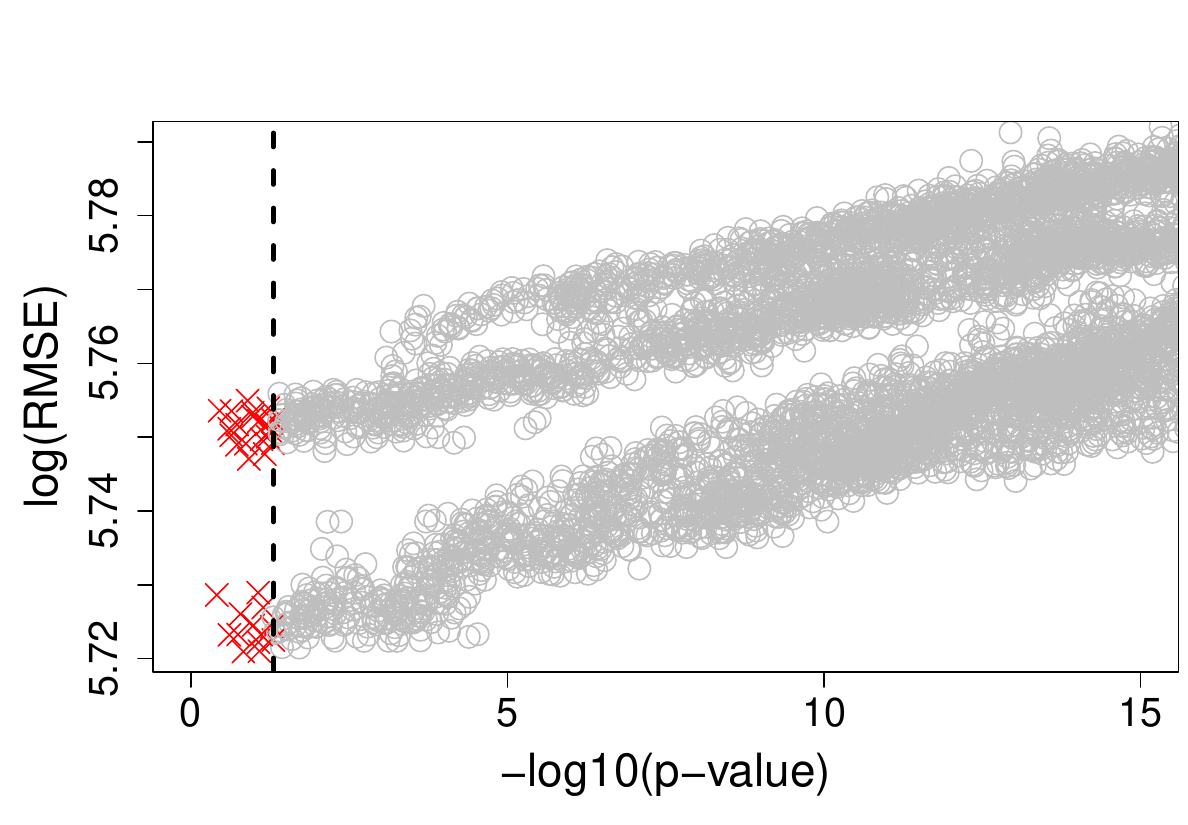}
    \caption{Scatterplot of one step ahead log(RMSE) versus $-\log_{10}(p\text{-value})$ of the LR  test between 28,672 linear time series models and a large benchmark model at hour 12 using the Italian electricity-load data. The vertical dashed line marks the $5\%$ significance threshold, separating rejected (gray circles) and not rejected (red crosses) models by the LR test.}
    \label{figure1}
\end{figure}

Our work is motivated by a specific empirical problem: modeling hourly electricity load in Italy over the period from 2019 to 2023. This type of data, augmented with temperature and renewable generation measures, naturally gives rise to many competing specifications, making model selection uncertainty unavoidable. Italy provides a  particularly informative case, as the sample includes major structural disruptions, particularly the COVID-19 pandemic and the 2022 energy crisis, that substantially affect load dynamics and complicate model choice.  As an illustration of selection uncertainty, Figure \ref{figure1} shows the logarithmic root mean squared  error  (RMSE) of 28,672 linear time series model specifications fitted with   electricity load data described in Section \ref{sec:empirics}. The $-\log_{10}(p\text{-value})$  from likelihood ratio (LR) tests comparing  a large benchmark model to all candidate  submodels are also reported. Notably, there are a number of models with similar prediction performance that cannot be discarded by the LR test. All these models provide statistically indistinguishable explanations of the true data generating process at a given level of accuracy.  


 Existing approaches to model selection typically focus on selecting a single optimal model. They can be broadly categorized into three classes: hypothesis testing \citep{diebold2002,Romano2005,Giacomini,CLARK2007291}, information criteria \citep{Billah01102005,EBIC,AIC,ding2018model}, and penalized methods \citep{tibshirani1996regression,fan2001variable,zou2005regularization,penalized,MCP,UematsuTanaka2019, Kock2020}.  
 Information criteria such as the Akaike Information Criterion (AIC) and the Bayesian Information Criterion (BIC) often lead to substantially different model choices, particularly in finite samples \citep{hansen2007least}, while regularization methods such as LASSO exhibit considerable selection variability in noisy or high-dimensional settings \citep{BuhlmannVanDeGeer2011HighDim,Kock2020}. When multiple models provide comparable explanatory power, conditioning inference on a single selected specification may overlook meaningful structural differences and understate the uncertainty inherent in the model selection process.

In this paper we address model selection uncertainty by introducing the Model Selection Confidence Set (MSCS) for univariate time
series models involving autoregressive and moving average components with exogenous inputs (ARMAX). The MSCS is defined as the set of all model specifications that are statistically indistinguishable from the true data-generating process at confidence level $(1-\alpha)\%$. Analogous to confidence intervals in parameter estimation, the MSCS size serves as a diagnostic for selection uncertainty: large sets reflect higher noise, with many models equally supported by the data while concentration around fewer models indicates lower uncertainty on the underlying data generating process.
A key feature of the MSCS is its subset of Lower Boundary Models (LBMs), i.e., the minimal models in the MSCS that cannot be further simplified without significant information loss. Studying LBMs helps identify key model terms: predictors and lags that appear across many LBMs are likely essential, whereas those that occur only sporadically may be less influential. In this way, the MSCS provides deeper  insight than traditional single-model selection or standard model-averaging techniques.

Several studies have been devoted in developing sets of models, each statistically justified by the data, to better account for model uncertainty. \cite{MCS} propose constructing a model confidence set (MCS) from a given set of candidate models by a sequence of equivalence tests on the currently remaining models, followed by a rule that removes the worst performers. Their method obtains a subset of the original models that is meant to contain the set of models with the best empirical performance under some given loss function. 
While sharing a similar motivation, our methodology builds on the confidence set framework of \cite{VSCS} for variable selection in linear models, which employs screening approaches based on the F-test, and later extended by \cite{Ferrari2019} and \cite{MCBounds}  to more general settings using the likelihood ratio test. The crucial distinction lies in their objective: the MCS seeks to identify the best predictive models from the candidate pool, whereas the latter framework aims to identify the true underlying process that generated the data.

This paper makes three main contributions. First, we develop a methodological framework for constructing confidence sets of univariate time series models, extending variable selection confidence set approaches to accommodate temporal dependencies in time series data. Second, we evaluate finite-sample performance through Monte Carlo experiments, confirming that the MSCS attains the desired asymptotic coverage probability and documenting how the sizes of the MSCS and the LBM vary with sample size and across different data-generating processes. Third, we apply the methodology to Italian hourly electricity load data, showing that accounting for model selection uncertainty substantially enhances forecasting robustness and reveals multiple plausible predictor combinations across different hours of the day.

The remainder of the paper is as follows. Section \ref{sec:methods} introduces the ARMAX modeling framework and develops the Model Selection Confidence Set (MSCS) methodology in this context. In the same section, we formally define Lower Boundary Models within the ARMAX framework. Section \ref{sec:mc} investigates the finite-sample behaviour of the procedure through Monte Carlo experiments under a range of data-generating processes and model misspecification scenarios. The analysis focuses on coverage probabilities, the size and composition of the MSCS, and the ability of the method to recover structurally relevant specifications. Section \ref{sec:empirics} applies the proposed methodology to hourly Italian electricity load data, examining both the empirical composition of the MSCS and the predictive performance of the models contained in the set. Particular attention is given to forecast accuracy and the stability of model terms inclusion across competing specifications. Section \ref{sec:conclusion} concludes with a discussion of the main findings and directions for future research.

\section{Model selection confidence set and lower boundary models} \label{sec:methods}

\subsection{Confidence set construction and asymptotic coverage}

Let $y_t$ be the electricity load at time $t= 1,...,T$. The autoregressive moving average model of order $(p,q)$ that incorporates $s$ exogenous variables observed with $r$ lags, denoted as ARMAX$(p,q,r)$, is defined as:

\begin{equation} \label{model}
y_{t} = \alpha + \sum_{j=1}^{p} \beta_j y_{t-j} + \sum_{k=1}^{q} \vartheta_k \varepsilon_{t-k} + \sum_{l=1}^{r} \boldsymbol{\eta}^\top_l \boldsymbol{x}_{t-l} + \varepsilon_t, \ \ t = 1, \dots, T,
\end{equation}
where  $x_t=(x_{t1}, \dots, x_{ts})^\top $, is a $s\times 1$ vector of predictors, $\varepsilon_1, \dots ,\varepsilon_t$ are independently and identically distributed  errors with zero mean and variance $\sigma^2$. Let $\boldsymbol{\theta }= (\alpha, \boldsymbol\beta^\top, \boldsymbol\vartheta^\top , \boldsymbol\eta^\top)^\top$ be the overall parameter vector where $\boldsymbol{\beta}= (\beta_1, \dots, \beta_p)^\top \in \mathbb{R}^p$, $\boldsymbol{\vartheta} = (\vartheta_1, \dots, \vartheta_q)^\top \in \mathbb{R}^q$ and $\boldsymbol{\eta}= (\eta^\top_1, \dots, \eta^\top_r)^\top \in \mathbb{R}^{sr}$ are the coefficients associated with lagged response variables, lagged error terms, and exogenous variables, respectively.  We consider  the situation where  multiple elements of $\boldsymbol{\theta}$ might be equal to zero. A model \( m \) is defined as a tuple \( m = \{S_{\text{AR}}, S_{\text{MA}}, S_{\text{X}}\} \), 
where \( S_{\text{AR}} = \{ j \in \{1, \dots, p\} \mid \beta_j \neq 0 \} \) represents the set 
of autoregressive lags with nonzero coefficients in the parameter vector  $\boldsymbol{\beta}$,  
\( S_{\text{MA}} = \{ k \in \{1, \dots, q\} \mid \vartheta_k \neq 0 \} \) is the set of 
moving average lags with nonzero coefficients in $\boldsymbol{\vartheta }$, and 
\( S_{\text{X}} = \{ (i, l) \mid i \in \{1, \dots, s\}, l \in \{1, \dots, r\}, \eta_{i,l} \neq 0 \} \) 
corresponds to the set of exogenous variables and their lags that have nonzero coefficients. The dimension of model $m$ is denoted by $d_m = |m|$; while the space of all feasible models is denoted by $\mathcal{M}$ and has cardinality 
\( |\mathcal{M}| = 2^{p + q + rs } \) , when all lags are considered. The true model is denoted by 
\( m^* = (S_{\text{AR}}^*, S_{\text{MA}}^*, S_{\text{X}}^*) \), where each subset contains 
the indices of terms with nonzero coefficients in the true data-generating process.

For a given confidence level $(1-\alpha)$, where $0 < \alpha < 1$ (e.g., 0.95 or 0.99), we aim to construct a confidence set of ARMAX models, $\widehat{\mathcal{M}}_\alpha$, satisfying

\begin{equation}\label{coverageP}
P(m^\ast \in \widehat{\mathcal{M}}_\alpha) \geq 1 - \alpha, \quad \text{as} \quad T \to \infty.
\end{equation}
This property, referred to as asymptotic coverage, ensures that the true data generating process  $m^\ast$ is included in the MSCS with large probability in large samples. In contrast to classical model selection procedures that return a single “best” specification, the MSCS acknowledges the intrinsic uncertainty of the selection step and delivers a set of models that are statistically indistinguishable at the prescribed confidence level. As a result, inference and prediction are no longer conditioned on a single selected model but instead accounts for the sampling variability inherent in model comparison. The coverage property therefore provides a formal frequentist guarantee that model uncertainty is explicitly controlled, rather than implicitly ignored.

In the context of ARMAX models, we rely on the likelihood ratio (LR) test to decide whether a given candidate model $m$ is statistically plausible compared to a full model, $m_f$, which includes all potentially useful predictors and lag terms.  The full model serves as a comprehensive benchmark, capturing the maximum possible structure in the data. In order for the LR test to be meaningful, we assume that $T> d_f = |m_f|$; then the candidate model $m$ is nested in $m_f$ and both models ($m, m_f$) are uniquely estimated via maximum likelihood under standard regularity conditions. Model $m$ is rejected when
\begin{equation} \label{eq:lrt}
  \Lambda( m) = 2\left( \log L(m_f) - \log L(m) \right) \geq q_{1-\alpha}(d_f-d_m),
\end{equation}
where $L(m)$ and $L(m_f)$ are  the maximized likelihood functions for models $m$ and $m_f$, respectively, and $q_{1-\alpha}(\delta)$ is the $(1-\alpha)$-quantile for the central chi-square random variable with $\delta$ degrees of freedom, denoted by $\chi^2_{\delta}$. The MSCS $\widehat{\mathcal{M}}_\alpha$ is defined by all the models not rejected by the LR test, i.e.,
\begin{equation}
\widehat{\mathcal{M}}_\alpha = \{ m: \Lambda(m)  < q_{1-\alpha}(d_f-d_m)\},
\end{equation}
and by default, the full model is included in $\widehat{\mathcal{M}}_\alpha$. By construction, if the true model is not the full model, the confidence set $\widehat{\mathcal{M}}_\alpha$ has asymptotic coverage probability of \( 1-\alpha \)  as $T \rightarrow \infty$, as stated in Eq. (\ref{coverageP}). 

This result follows trivially from the fact that when \( m = m^{\ast} \),  under the null hypothesis that the restricted model \( m \) is correctly specified, we have that \( \Lambda(m) \xrightarrow{d} \chi^2_{d_f - d_m} \), as \( T \to \infty \). A proof for  general likelihoods and diverging number of parameters is available in Theorem 1 of \cite{Ferrari2019}. Although we do not provide formal results in this paper, following similar steps one can show that the same behavior holds in the current setting for fixed number of parameters, $d < T$, under standard regularity conditions for the LR test in ARMAX models involving: (i) the correct specification of the true model, (ii) identifiability of parameters, (iii) finite and nonsingular Fisher information matrix, (iv) weak dependence conditions such as strong mixing or ergodicity, and  (v) regularity of the innovation process, where the error terms \( \varepsilon_t \) are assumed to be i.i.d. or weakly dependent with finite variance.  Our procedure works also when  alternative tests, such as the Wald or the Lagrange multiplier (score) tests (e.g., see \cite{WaldLR}), are employed for model screening since they are asymptotically equivalent under standard regularity conditions. 

The confidence set \( \widehat{\mathcal{M}}_\alpha \) allows us to determine whether a model \( m \) is overly parsimonious, meaning that it may exclude important variables. If \( m \notin \widehat{\mathcal{M}}_\alpha \), such a model likely overlooks some key terms. Unlike traditional model selection methods that force a single best model, the MSCS explicitly quantifies selection uncertainty by identifying a set of plausible models. This prevents overconfidence in a single specification and mitigates the risk of misleading conclusions when multiple models fit the data similarly well. Moreover, the structure of \( \widehat{\mathcal{M}}_\alpha \) provides insights into predictor stability, distinguishing between core explanatory variables and those whose inclusion is selection-dependent. The size of the MSCS is also informative: a small confidence set suggests strong support for a particular specification, while a larger one signals greater uncertainty, revealing multiple competing models.

 \subsection{Lower boundary models}\label{sec:lbm}

The MSCS described in the previous section could have a large size mainly due to the possibility of including irrelevant parameters that survive the likelihood ratio screening. To manage the potential size of the MSCS, we focus on a smaller, yet  informative subset, the set of lower boundary models (LBMs), denoted by $\widehat{\mathcal{B}}_\alpha$, defined by the set of models in  $\widehat{\mathcal{M}}_\alpha$ without any nested sub-models in $\widehat{\mathcal{M}}_\alpha$. Intuitively, the LBMs is the collection of the most parsimonious models that remain well-supported by the data at a given confidence level.

\cite{VSCS} study the LBMs in the context of linear models finding that both their composition and size of $\widehat{\mathcal{M}}_\alpha$   provide valuable insights into overall variable selection uncertainty. Their findings on LBMs can be extended to the ARMAX setting when the total number of parameters in the full model, $d$, is fixed, and the sample size, $T$, is large under the following  scenarios.

\begin{itemize}
    \item  {\it No model selection uncertainty.} $\widehat{\mathcal{B}}_\alpha$ contains a single model suggesting that all terms included in the ARMAX model are essential, with no redundant terms. As the sample size increases, $\widehat{\mathcal{B}}_\alpha$ converges to a set containing the single true model, reflecting an increasingly complete understanding of the data-generating process.
    \item {\it Moderate model selection uncertainty.} A relatively small $\widehat{\mathcal{B}}_\alpha$ indicates slight variations in model selection, with core explanatory variables being consistently included, while a few terms remain uncertain. The presence of multiple valid models signals model selection ambiguity.
    \item {\it Pronounced model selection uncertainty.} A large $\widehat{\mathcal{B}}_\alpha$ highlights the difficulty in pinpointing a single best model. Thus, model selection criteria may yield multiple near-optimal solutions, illustrating the inherent complexity of the selection process.
\end{itemize}
 As \( T \to \infty \), the lower boundary models increasingly    include the terms of the true data-generating process with increasing probability, effectively identifying all relevant model terms. Particularly, as \( T \to \infty \), if \( m^\ast \) is not the full model, then $P(\widehat{\mathcal{B}}_\alpha = \{m^\ast\}) \to 1 - \alpha$, see  \cite{VSCS} for this result in linear models. A similar result may be established for ARMAX processes by following  Theorem 2 of \cite{Ferrari2019}, provided that standard regularity conditions hold to ensure the asymptotic validity of the likelihood ratio test.

In this paper, we observe that the cardinality and the composition of LBMs provide valuable insights into model terms selection uncertainty in time-dependent settings. Furthermore, by examining how these lower boundary models relate to more complex models in the MSCS, researchers can gain insights into the essential versus potentially superfluous terms, thus offering a principled approach to understand the trade-off between model simplicity and explanatory power.

The finite-sample behavior of $\widehat{\mathcal{B}}_\alpha$ is evaluated in Section~\ref{sec:mc} via Monte Carlo experiments, while in Section~\ref{sec:empirics} we illustrate how the cardinality and composition of LBMs inform model selection uncertainty in electricity load forecasting.

Moreover, to reduce the risk of being excessively parsimonious by omitting important terms that might appear in only some  models we explore the utility of a combination of LBMs, called union model. This specification is defined as $m^{\text{max}} = \cup_{m \in \widehat{\mathcal{B}_\alpha}}  m$. The union model is especially valuable when dealing with noisy temporal data: while individual models might capture specific patterns, the union model $m^{\text{max}}$ ensures all relevant temporal dynamics are represented. This is confirmed in Monte Carlo experiments in Section \ref{sec:mc} showing that $m^{\text{max}}$ consistently contains the true model across all settings, with a probability of nearly 1,
underscoring its robustness in covering the correct model, even in small samples.

\subsection{Inclusion importance metrics}

To quantify the relevance of specific terms, we define the inclusion importance metrics, denoted as \(  {II}(\theta_j) \), defined as the frequency with which a given parameter $ \theta_j  \in \Theta$
 appears as nonzero in the LBMs of the selected ARMAX confidence set. This metric provides insight into the relevance and stability  of individual predictors across competing models and it is defined as
\begin{equation}
II(\theta_j) = \dfrac{K(\theta_j)}{ | \widehat{\mathcal{B}}_\alpha | }, \quad \theta_j  \in \Theta ,
\end{equation}
where $K(\theta_j)$ is the number of times  $\theta_j$ appears in $\widehat{\mathcal{B}}_\alpha $ and \( | \widehat{\mathcal{B}}_\alpha  | \), represents the cardinality of the lower boundary $\widehat{\mathcal{B}}_\alpha $. In the context of ARMAX modeling, the inclusion importance determines which exogenous variables and lagged terms are most important when sampling variability is considered. A large value of \( {II}(\theta_j) \) indicates that $\theta_j$ is a core model term, whereas a small value suggests that $\theta_j$ has limited importance.

We extend the inclusion importance measure to the entire confidence set, \( \widehat{\mathcal{M}}_\alpha \). Specifically, we define the inclusion importance for the entire MSCS as $
\widetilde{II}(\theta_j) = {\widetilde{K}(\theta_j)}/{|\widehat{\mathcal{M}}_\alpha|}$, where \( \widetilde{K}(\theta_j) \) denotes the number of times the parameter \( \theta_j \) is nonzero across all models in \( \widehat{\mathcal{M}}_\alpha \). Since the LR test tends to retain models that are larger than the true model, irrelevant predictors appear randomly in approximately half of the models within \( \widehat{\mathcal{M}}_\alpha \). To account for this behavior, we define a normalized inclusion importance metric that corrects for the expected inclusion rate of spurious predictors:
\begin{equation}\label{eq:normalized_II}
\widetilde{II}_s(\theta_j) = \max\left\{0, \frac{\widetilde{II}(\theta_j) - 0.5}{0.5} \right\}.
\end{equation}
The normalized inclusion importance emphasizes the terms with strong inclusion signals while down-weights those that appear inconsistently due to overfitting. A large value of $\widetilde{II}_s(\theta_j)$ indicates that a predictor is included in all models, while a value near zero suggests weak or spurious inclusion.

\section{Monte Carlo experiments} \label{sec:mc}

To study the finite-sample properties of the MSCS and LBMs, we conduct several Monte Carlo experiments by generating data from  six ARMAX models specified by various configurations of the autoregressive, moving average and exogenous components as reported in Table \ref{tab:gdps}.  For all the models, we set $\sigma =0.8$  and consider both independent and correlated predictors generated from a multivariate normal distribution taking,  uniform correlation  between each predictor pair with values $\rho \in \{0, 0.7\}$. All the considered models do not include the intercept,  specify a zero-lag for the exogenous variables and have normally distributed error terms.  In each experiment, we consider 500 Monte Carlo samples of size $T \in \{ 250, 1000, 10000\}$. The MSCS is computed using exhaustive enumeration of the sample space consisting of all feasible submodels of the full model ARMAX($3$, $2$, $6$) and  consider nominal confidence levels $(1-\alpha) \in \{0.95,0.99\}$.

\begin{table}[h]    
\centering
    \begin{tabular}{cccccccccccc}
        Model & ($p$, $q$, $x$) & & $\beta_1$ & $\beta_2$ & $\vartheta_1$ & $\eta_1$ & $\eta_2$ & $\eta_3$ & $\eta_4$ & $\eta_5$ & $\eta_6$ \\
        \hline
        A & ($1$, $0$, $2$)       && 0.8  &  0 &  0  & -2.0 &0 & 2.0   & 0&0&0 \\
        B& ($1$, $0$, $5$)         && 0.8  & 0 & 0  & -2.0&0  & 2.0    & 1&1.5&-1.5 \\
        C &($1$, $1$, $2$)      && 0.7  &  0 & 0.5 & -2.0 &0 & 2.0   &0&0&0 \\
        D& ($1$, $1$, $5$)     && 0.7  &  0 & 0.5& -2.0 &0 & 2.0    & 1.0&1.5&-1.5 \\
        E& ($2$, $1$, $2$)     && 0.6  & -0.2  & 0.5  & -2.0 &0  & 2.0   &0 &0 &0 \\
        F & ($2$, $1$, $5$) && 0.6  & -0.2  & 0.5  & -2.0 &0 & 2.0    & 1.0&1.5&-1.5 \\
    \end{tabular}
    \caption{Parameter values for different data generating processes (A--F) considered in the Monte Carlo experiments where $\beta_j$, $\vartheta_k$ and $\eta_l$ refer to the AR, MA and exogenous components, respectively.}
    \label{tab:gdps}
\end{table}

For each setting, we study the key properties of the MSCS and related LBMs in finite samples. Particularly, we compute Monte Carlo estimates of the following quantities: 1) MSCS size $|\widehat{\mathcal{M}}_{\alpha}|$; 2) average size of models in the lower boundary $\widehat{\mathcal{B}}_{\alpha}$; 3) coverage probability  $P_\alpha^* = P(m^* \in \widehat{\mathcal{M}}_{\alpha})$; 4)   expected distance $ d^{*, max} = E[d(m^\ast, m^{\text{max}})]$, where $d(m_1, m_2)$ is the Hamming distance counting the number of different elements in $m_1$  compared to model $m_2$; 5) probability that the true model is included in the LBM union model, $P_\alpha^{max} = P(m^\ast \subseteq m^{\text{max}})$, where $m^{\text{max}}$ is defined in Section \ref{sec:lbm}.

\begin{table}[htbp]
\centering
\label{tab:vscs_lbm_size}
\begin{tabular}{llcc|rr|rr}
\toprule
& & & & \multicolumn{2}{c}{$x=2$} & \multicolumn{2}{c}{$x=5$} \\
\cmidrule(lr){5-6} \cmidrule(lr){7-8}
Model & $T$ & $(1-\alpha)$\% & $\rho$ & $|\mathcal{M}_{\alpha}|$ & $|\mathcal{B}_{\alpha}|$ & $|\mathcal{M}_{\alpha}|$ & $|\mathcal{B}_{\alpha}|$ \\
\midrule
A-B & 250 & 99\% & 0 & 128.6 (1.4) & 1.1 (0.0) & 14.7 (0.2) & 1.0 (0.0) \\
          &     &       & 0.7 & 128.3 (1.5) & 1.1 (0.0) & 14.6 (0.2) & 1.0 (0.0) \\
          & 1000 & 99\% & 0 & 131.4 (1.3) & 1.0 (0.0) & 15.1 (0.2) & 1.0 (0.0) \\
          &      &       & 0.7 & 131.2 (1.3) & 1.0 (0.0) & 15.1 (0.2) & 1.0 (0.0) \\
          & 10000 & 99\% & 0 & 141.2 (0.4) & 1.0 (0.0) & 16.7 (0.1) & 1.0 (0.0) \\
          &       &       & 0.7 & 141.3 (0.4) & 1.0 (0.0) & 16.7 (0.1) & 1.0 (0.0) \\
          
        & 250 & 95\% & 0 & 104.1 (2.1) & 1.2 (0.0) & 11.2 (0.3) & 1.1 (0.0) \\
          &     &       & 0.7 & 104.4 (2.1) & 1.2 (0.0) & 11.2 (0.3) & 1.1 (0.0) \\
          & 1000 & 95\% & 0 & 112.6 (2.0) & 1.2 (0.0) & 12.5 (0.3) & 1.1 (0.0) \\
          &      &       & 0.7 & 112.4 (2.0) & 1.2 (0.0) & 12.6 (0.3) & 1.1 (0.0) \\
          & 10000 & 95\% & 0 & 134.6 (0.9) & 1.1 (0.0) & 15.8 (0.1) & 1.0 (0.0) \\
          &       &       & 0.7 & 134.7 (0.9) & 1.1 (0.0) & 15.8 (0.1) & 1.0 (0.0) \\
\midrule
C-D & 250 & 99\% & 0 & 114.9 (0.8) & 2.0 (0.0) & 13.3 (0.1) & 1.9 (0.0) \\
          &     &       & 0.7 & 114.5 (0.9) & 2.1 (0.0) & 13.3 (0.1) & 1.9 (0.0) \\
          & 1000 & 99\% & 0 & 101.0 (0.5) & 1.7 (0.0) & 11.6 (0.1) & 1.5 (0.0) \\
          &      &       & 0.7 & 101.2 (0.5) & 1.7 (0.0) & 11.6 (0.1) & 1.5 (0.0) \\
          & 10000 & 99\% & 0 & 93.9 (0.3) & 1.0 (0.0) & 10.9 (0.0) & 1.0 (0.0) \\
          &       &       & 0.7 & 93.7 (0.3) & 1.0 (0.0) & 10.8 (0.0) & 1.0 (0.0) \\
          & 250 & 95\% & 0 & 102.7 (1.2) & 2.1 (0.0) & 11.8 (0.1) & 1.8 (0.0) \\
          &     &       & 0.7 & 102.8 (1.2) & 2.0 (0.0) & 11.9 (0.1) & 1.8 (0.0) \\
          & 1000 & 95\% & 0 & 92.4 (0.9) & 1.5 (0.0) & 10.6 (0.1) & 1.3 (0.0) \\
          &      &       & 0.7 & 93.0 (0.8) & 1.5 (0.0) & 10.6 (0.1) & 1.3 (0.0) \\
          & 10000 & 95\% & 0 & 90.2 (0.6) & 1.1 (0.0) & 10.4 (0.1) & 1.0 (0.0) \\
          &       &       & 0.7 & 90.1 (0.6) & 1.0 (0.0) & 10.4 (0.1) & 1.0 (0.0) \\
\midrule
E-F & 250 & 99\% & 0 & 106.2 (0.8) & 2.1 (0.0) & 12.0 (0.1) & 1.9 (0.0) \\
          &     &       & 0.7 & 106.1 (0.8) & 2.2 (0.0) & 12.0 (0.1) & 1.9 (0.0) \\
          & 1000 & 99\% & 0 & 83.5 (0.5) & 2.2 (0.0) & 9.3 (0.1) & 1.9 (0.0) \\
          &      &       & 0.7 & 83.5 (0.5) & 2.2 (0.0) & 9.3 (0.1) & 1.9 (0.0) \\
          & 10000 & 99\% & 0 & 63.0 (0.2) & 1.1 (0.0) & 7.0 (0.0) & 1.0 (0.0) \\
          &       &       & 0.7 & 62.8 (0.2) & 1.1 (0.0) & 7.0 (0.0) & 1.0 (0.0) \\
          & 250 & 95\% & 0 & 91.2 (1.1) & 2.1 (0.0) & 10.2 (0.1) & 1.8 (0.0) \\
          &     &       & 0.7 & 90.9 (1.1) & 2.2 (0.0) & 10.3 (0.1) & 1.8 (0.0) \\
          & 1000 & 95\% & 0 & 74.0 (0.7) & 2.0 (0.0) & 8.2 (0.1) & 1.8 (0.0) \\
          &      &       & 0.7 & 73.9 (0.7) & 2.1 (0.0) & 8.2 (0.1) & 1.8 (0.0) \\
          & 10000 & 95\% & 0 & 60.7 (0.3) & 1.1 (0.0) & 6.7 (0.0) & 1.0 (0.0) \\
          &       &       & 0.7 & 60.4 (0.4) & 1.1 (0.0) & 6.8 (0.0) & 1.0 (0.0) \\
\bottomrule
\end{tabular}
\caption{Monte Carlo estimates for the MSCS size ($|\mathcal{M}_{\alpha}|$) and LBM set size ($|\mathcal{B}_{\alpha}|$) based on 500 samples of size $T \in \{ 250, 1000, 10000\}$ for models A--F described in Table~\ref{tab:gdps}. Monte Carlo standard errors are reported in parenthesis.}\label{MC_size}
\end{table}

Table \ref{MC_size} shows Monte Carlo estimates for the MSCS  and LBM set size. The number of LBMs is relatively small, suggesting that even when the MSCS contains many models, only a few of them are  essential for capturing the underlying process. The presence of correlated exogenous predictors (\(\rho = 0.7\)) does not significantly alter these trends, indicating that the selection procedure remains robust under moderate correlation.

When comparing different confidence levels, both LBM set and MSCS size increase with the confidence level, given the stricter threshold for model rejection. 

The behavior of the MSCS and LBM set exhibits distinct dependencies on sample size. The LBM set size consistently decreases as the sample size increases, converging to 1, as expected, but the rate of convergence varies depending on model complexity. For the ARMA(1,0) model (A--B), the LBM size remains close to 1 even in small samples, indicating precise model selection. In contrast, the ARMA(1,1) model (C--D) initially shows larger LBM sizes of about $2$  due to greater uncertainty in small samples caused by estimation fluctuations but gradually converges to 1 as the sample size increases. The ARMA(2,1) case (E--F)  exhibits the slowest convergence, with an initial over-selection of parameters ranging from about $2.1$ to $2.2$, as the model selection process struggles with the interaction of autoregressive and moving average components. The presence of MA terms introduces greater model uncertainty, delaying MSCS stabilization. However, as the sample size grows, estimation variance decreases, leading to more accurate model identification, and the LBM size approaches 1. 

The behavior of the full MSCS is less consistent and depends on whether the true model is ARMA(1,1) or ARMA(1,0). Since ARMA(1,0) is structurally farther from the full ARMA(3,2) model, more models survive the LR test, making rejection less frequent. In small samples, the LR test struggles to distinguish between an AR(1) model with short-term noise and an ARMA(1,1) model, leading to frequent misclassification. With increasing sample size, reduced estimation variance allows for a clearer identification of the MA(1) component, shrinking the MSCS. Conversely, when the true model is ARMA(1,0), small samples make it difficult to detect spurious MA components, resulting in a more conservative selection. As sample size grows, small but systematic autocorrelations in the residuals may mimic a MA structure, increasing the likelihood of selecting models with unnecessary MA terms, thereby expanding the MSCS.

\begin{table}[htbp]
\centering
\setlength{\tabcolsep}{4pt}
\renewcommand{\arraystretch}{1.1}

{\normalsize
\begin{tabular}{@{}>{\normalsize}l *{3}{>{\normalsize}c}|*{3}{>{\normalsize}c}|*{3}{>{\normalsize}c}@{}}
\toprule
& & & & \multicolumn{3}{c|}{$x=2$} & \multicolumn{3}{c}{$x=5$} \\
\cmidrule(lr){5-7} \cmidrule(lr){8-10}
Model & $T$ & $(1-\alpha)\%$ & $\rho$ &
$P_\alpha^*$ & $d^{*,\max}$ & $P_\alpha^{\max}$ &
$P_\alpha^*$ & $d^{*,\max}$ & $P_\alpha^{\max}$ \\
\midrule
A-B & 250   & 99\% & 0   & 0.92 & 0.34 & 1.00 & 0.92 & 0.25 & 1.00 \\
  &       &      & 0.7 & 0.91 & 0.34 & 1.00 & 0.89 & 0.32 & 1.00 \\
  & 1000  & 99\% & 0   & 0.95 & 0.16 & 1.00 & 0.94 & 0.19 & 1.00 \\
  &       &      & 0.7 & 0.95 & 0.18 & 1.00 & 0.93 & 0.21 & 1.00 \\
  & 10000 & 99\% & 0   & 0.99 & 0.00 & 1.00 & 0.99 & 0.01 & 1.00 \\
  &       &      & 0.7 & 0.99 & 0.03 & 1.00 & 0.99 & 0.01 & 1.00 \\
 & 250   & 95\% & 0   & 0.81 & 0.82 & 1.00 & 0.74 & 0.80 & 1.00 \\
  &       &      & 0.7 & 0.80 & 0.84 & 1.00 & 0.75 & 0.80 & 1.00 \\
  & 1000  & 95\% & 0   & 0.84 & 0.65 & 1.00 & 0.78 & 0.74 & 1.00 \\
  &       &      & 0.7 & 0.83 & 0.72 & 1.00 & 0.78 & 0.64 & 1.00 \\
  & 10000 & 95\% & 0   & 0.96 & 0.15 & 1.00 & 0.96 & 0.08 & 1.00 \\
  &       &      & 0.7 & 0.96 & 0.16 & 1.00 & 0.95 & 0.12 & 1.00 \\
\midrule
C-D & 250   & 99\% & 0   & 0.97 & 1.40 & 1.00 & 0.95 & 1.32 & 0.99 \\
  &       &      & 0.7 & 0.97 & 1.45 & 1.00 & 0.96 & 1.34 & 0.99 \\
  & 1000  & 99\% & 0   & 0.97 & 1.36 & 1.00 & 0.98 & 0.97 & 1.00 \\
  &       &      & 0.7 & 0.98 & 1.36 & 1.00 & 0.98 & 0.95 & 1.00 \\
  & 10000 & 99\% & 0   & 0.99 & 0.01 & 1.00 & 0.98 & 0.03 & 1.00 \\
  &       &      & 0.7 & 0.99 & 0.01 & 1.00 & 0.98 & 0.04 & 1.00 \\
 & 250   & 95\% & 0   & 0.89 & 1.76 & 0.99 & 0.89 & 1.42 & 0.99 \\
  &       &      & 0.7 & 0.89 & 1.71 & 0.99 & 0.90 & 1.37 & 0.99 \\
  & 1000  & 95\% & 0   & 0.93 & 1.15 & 1.00 & 0.91 & 0.70 & 0.99 \\
  &       &      & 0.7 & 0.93 & 1.17 & 1.00 & 0.91 & 0.68 & 0.99 \\
  & 10000 & 95\% & 0   & 0.95 & 0.13 & 1.00 & 0.95 & 0.09 & 1.00 \\
  &       &      & 0.7 & 0.95 & 0.11 & 1.00 & 0.95 & 0.10 & 1.00 \\
\midrule
E-F & 250   & 99\% & 0   & 0.97 & 0.93 & 0.91 & 0.97 & 0.99 & 0.82 \\
  &       &      & 0.7 & 0.97 & 1.00 & 0.92 & 0.97 & 0.97 & 0.83 \\
  & 1000  & 99\% & 0   & 0.98 & 1.46 & 0.79 & 0.98 & 1.14 & 0.83 \\
  &       &      & 0.7 & 0.98 & 1.46 & 0.80 & 0.98 & 1.14 & 0.82 \\
  & 10000 & 99\% & 0   & 0.99 & 0.08 & 1.00 & 0.99 & 0.04 & 1.00 \\
  &       &      & 0.7 & 0.99 & 0.09 & 1.00 & 0.99 & 0.04 & 1.00 \\
 & 250   & 95\% & 0   & 0.92 & 1.35 & 0.84 & 0.90 & 1.20 & 0.73 \\
  &       &      & 0.7 & 0.90 & 1.46 & 0.85 & 0.90 & 1.16 & 0.73 \\
  & 1000  & 95\% & 0   & 0.94 & 1.34 & 0.88 & 0.94 & 0.92 & 0.92 \\
  &       &      & 0.7 & 0.94 & 1.38 & 0.89 & 0.94 & 0.92 & 0.92 \\
  & 10000 & 95\% & 0   & 0.96 & 0.12 & 1.00 & 0.95 & 0.07 & 1.00 \\
  &       &      & 0.7 & 0.95 & 0.16 & 1.00 & 0.95 & 0.07 & 1.00 \\
\bottomrule
\end{tabular}
}
\caption{Monte Carlo estimates for coverage probability of the MSCS ($P_\alpha^*$), average Hamming distance between the union of LBMs and the true model ($P_\alpha^{\max}$), and probability that the union of LBMs contains the true model ($P_\alpha^{\max}$), based on 500 samples of size $T \in \{250, 1000, 10000\}$ for models A--F described in Table~\ref{tab:gdps}. Monte Carlo standard errors are smaller than 0.01 for columns 5, 7, 8, and 10 and smaller than 0.1 for columns 6 and 9.}
\label{coverage_distance}
\end{table}

Table \ref{coverage_distance} shows estimates of the coverage probability of the MSCS and the distance of the true model from the LBMs. While the coverage approaches the nominal level with large $T$, in small samples it is not the case due to the finite-sample behavior of the likelihood ratio (LR) test, which tends to over-reject simpler models when the true model is nested within a larger one. Thus, in relatively small samples the LR screening somewhat struggles to distinguish between alternative specifications due to increased estimation variability and bias. The Hamming distance between $m^{\text{max}}$ and $m^\ast$ displays high values in small samples, thus indicating pronounced model selection uncertainty, but decreases sharply with increasing sample size.  Notably, the union of LBMs consistently contains the true model across all settings, with a probability of nearly 1, underscoring the robustness of the LBM approach in covering the correct model, even in small samples. These results confirm that while small-sample variability introduces uncertainty in model selection, the union of LBMs provides a safe model that ensures the presence of the true term, and increasing the sample size leads to more accurate model identification.

\section{Model selection and uncertainty in electricity load forecasting}\label{sec:empirics}

Recent empirical work on electricity load forecasting lacks consensus on a comprehensive set of load drivers. Temperature and calendar effects are the most acknowledged determinants \citep{bashiri2023}, with load responding nonlinearly to temperature \citep{henley1997non, damm2017effects}. Given this uncertainty, we adopt the MSCS methodology to assess the model uncertainty and study the importance of individual model terms.

\begin{figure}
    \centering
    \includegraphics[width=1\linewidth]{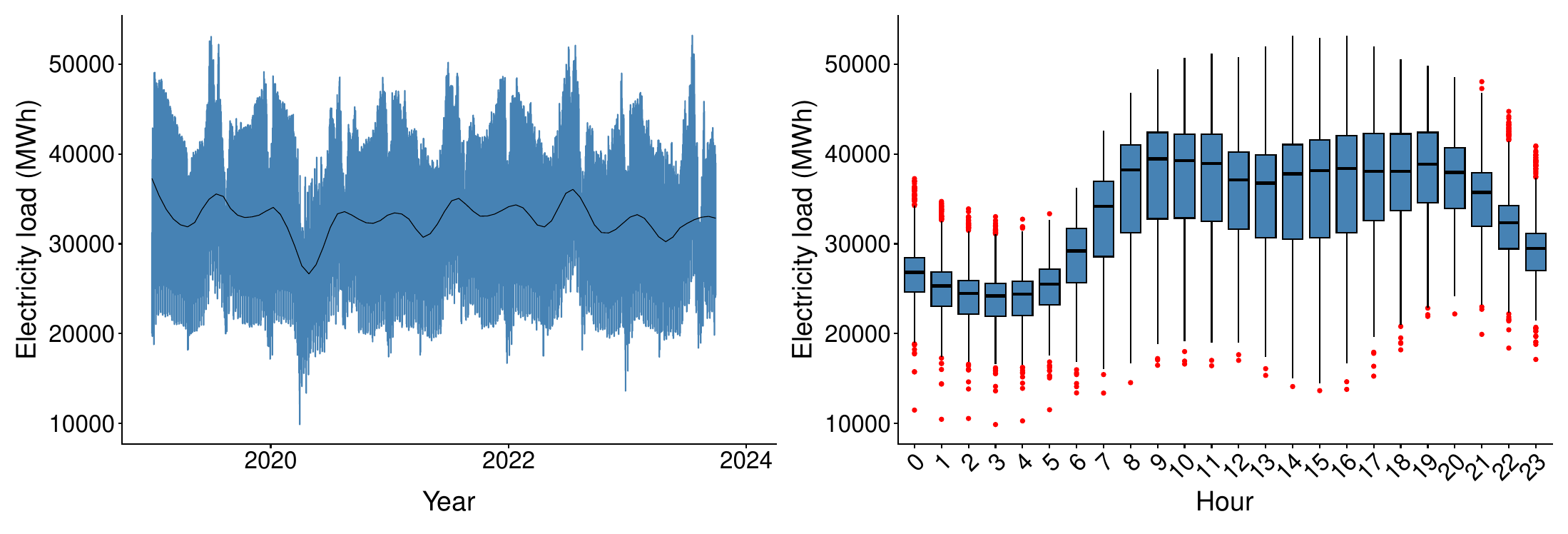}
    \caption{Raw hourly electricity load data with smoothed trend line (left) and boxplots for data grouped by hour of the day (right) for Italy from January 2019 to September 2023.}
    \label{ts}
\end{figure}

We analyze hourly electricity load data (MWh) for Italy obtained from the London Stock Exchange Group (LSEG) (\url{https://www.lseg.com/en/data-analytics}), from January 1 2019 to September 29 2023 consisting of a total of 41592 observations. Figure \ref{ts} shows the hourly load (MWh) time series for Italy, which exhibits moderate annual seasonality, with peaks during winter and summer due to increased heating and cooling loads. We collect a set of key predictors to explain variations in electricity load; these include forecasted solar (\texttt{Solar}) and wind (\texttt{Wind}) generation in MWh as renewable energy sources, as well as temperature (\texttt{Temperature}) and its squared term (\texttt{Temperature$^2$}) to capture nonlinear effects on load \footnote{Hourly temperature data were retrieved from the Milan Lambrate meteorological station through the Lombardy Regional Agency for Environmental Protection (\href{https://www.arpalombardia.it/temi-ambientali/meteo-e-clima/form-richiesta-dati/}{ARPA Lombardia}), specifically form sensor ID 2001.}. Calendar effects are modeled through indicator variables for months from December to February (\texttt{Winter}) and from June to August (\texttt{Summer}), for weekends and public holidays (\texttt{Week/Hol.}), and for the first COVID-19 lockdown in Italy (\texttt{Lockdown}), which had a noticeable impact on electricity consumption patterns. To account for the inherent persistence in load, we consider intra-day lagged values of electricity load of one (\texttt{Lag$_{1}$}), two (\texttt{Lag$_{2}$}) and three (\texttt{Lag$_{3}$}) hours.

\subsection{Quantifying selection uncertainty and key predictor relevance}

To construct the MSCS, we compare the full ARMAX($7$,$2$) model incorporating 11 exogenous predictors detailed in Section \ref{sec:empirics}  against all possible submodels \footnote{To work on a manageable model space we do not allow for gaps in auto-regressive and moving-average orders and we do not consider lagged values of external predictors. Cardinality of $\mathcal{M}$ is then restricted to 28672 different ARMAX structures.} based on the LR statistic as defined in Equation (\ref{eq:lrt}).  Setting $p=7$ ensures coverage of weekly consumption cycles, capturing recurring electricity load at the same hour across different days. Similarly, $q=2$ provides sufficient coverage of short-term autocorrelations in forecasting errors, particularly those influenced by daily fluctuations. Given the hourly nature of the data, we split the dataset into 24 separate time series, each corresponding to a specific hour of the day, and focus mainly on working hours (8-17).

\begin{table}[htbp]
\centering
\begin{adjustbox}{width=\textwidth,keepaspectratio}
\begin{tabularx}{\textwidth}{ll*{9}{>{\centering\arraybackslash}X}}
\toprule
\multicolumn{2}{c}{} & \multicolumn{9}{c}{\textbf{Hour}} \\
\cmidrule(lr){3-11}
&  & 8 & 9 & 10 & 11 & 12 & 13 & 14 & 15 & 16 \\
\midrule
\multirow{2}{*}{$|\mathcal{M}_{\alpha}|$} & $\alpha$ = 0.01 & 126 &  9 & 97 & 15 & 127 & 33 & 121 & 3029 & 249 \\
                          & $\alpha$ = 0.05 &  88 &  4 & 44 & 14 &  33 & 18 &  80 & 2647 & 131 \\
\multirow{2}{*}{$|\mathcal{B}_{\alpha}|$} & $\alpha$ = 0.01 &   1 &  2 &  1 &  1 &   3 &  3 &   5 &   3 &  3 \\
                          & $\alpha$ = 0.05 &   3 &  2 &  1 &  1 &   3 &  2 &   5 &   4 &  3 \\
\bottomrule
\end{tabularx}
\end{adjustbox}
\caption{MSCS size ($|\mathcal{M}_{\alpha}|$)  and LBMs size ($|\mathcal{B}_{\alpha}|$) across working hours for Italian electricity load data at 99\% and 95\%  confidence level.}
\label{ECS_LBM_size}
\end{table}

In Table \ref{ECS_LBM_size}, we present the size of the MSCS and its corresponding LBMs set, revealing distinct patterns in model selection uncertainty across working hours. The MSCS size remains consistently low throughout the morning and early afternoon, suggesting moderate uncertainty. This stability ends at hour 15, when the MSCS size increases by several orders of magnitude, indicating a rise in uncertainty, and then vanishing the hour after. 

The LBM set size  remains small across all hours, yet exhibits meaningful variation. In the morning, the set size is minimal, suggesting only moderate uncertainty in identifying the simplest adequate model. Moving into the afternoon, the set size increases, corresponding to increased uncertainty about which minimal specification best captures load dynamics. Crucially, this expansion in LBM diversity peaks at hour 14, one hour before the explosion in MSCS size. This temporal sequence, combined with a substantial reduction in the model's autoregressive order $p$ around the same time, provides evidence of a significant structural shift in the underlying uncertainty pertaining the data-generating process during the afternoon hours.

\begin{figure}
    \centering
    \includegraphics[width=1\linewidth]{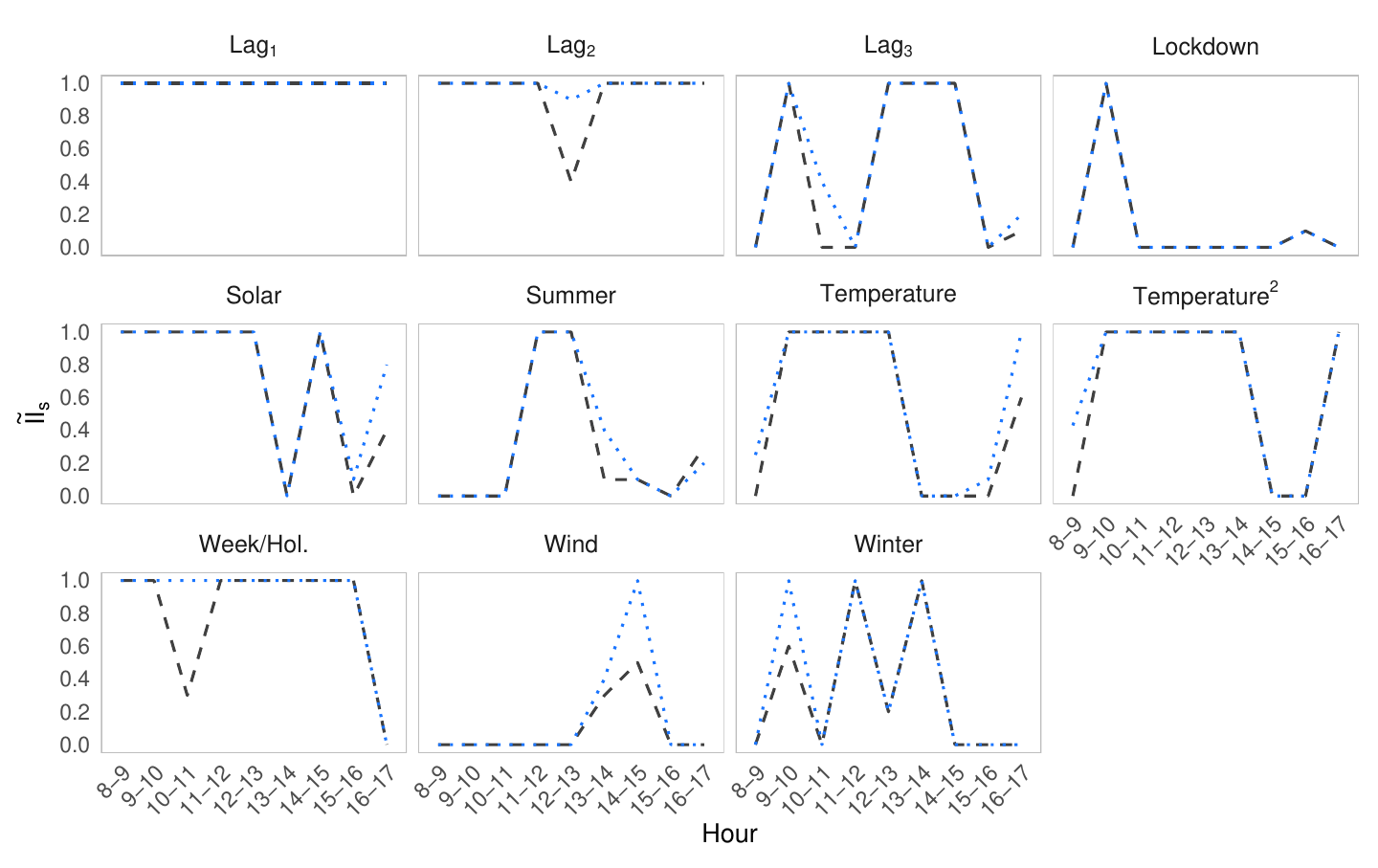}
    \caption{Normalized inclusion importance ($\widetilde{II_s}$) of different predictors at 95\%
(blue dotted) and 99\% (black dashed) confidence level for working hours.}
    \label{IIs}
\end{figure}

Table \ref{LBMs} presents the inclusion importance of different model terms of the lower boundary models corresponding to the MSCS for working hours. Despite the varying uncertainty reflected in LBM set sizes, the core explanatory variables  such as \texttt{Week/Hol.}, \texttt{Temperature}, \texttt{Temperature$^2$} and \texttt{Solar} and key lags like \texttt{Lag$_{1}$}, \texttt{Lag$_{2}$} remain largely consistent across all hours. Consistent selection of such predictors confirms the dominant influence of intraday  hourly lags, calendar effects, and weather conditions on electricity load.

Further insights emerge when comparing the structures of LBMs selected at the 99\% and 95\% confidence levels.  For instance, the 99\% confidence level produces greater diversity in the model's parameters specification, admitting various autoregressive structures across the boundary models. Conversely, for the same hours, the 95\% confidence level often results in a consensus on a single, specific structure, indicating a clearer model preference.

Beyond these core terms, the inclusion of environmental and event-based predictors reveals more nuanced patterns in electricity load. \texttt{Solar} appears in at least one boundary model for all daylight hours, aligning with its direct impact on the electricity grid through photovoltaic generation and its correlation with cooling load. \texttt{Wind}, however, is included far less frequently, suggesting its influence is more sporadic or less significant than other factors. The seasonal and event variables---\texttt{Summer}, \texttt{Winter} and \texttt{Lockdown}---also appear in specific LBMs rather than forming a consistent pattern, which implies their impact is concentrated in certain hours of the day rather than exerting uniform effects.

The autoregressive and moving average components exhibit a structured but highly flexible pattern. While many hours favor a weekly autoregressive dependency with $p=7$, reflecting the strong weekly cyclicality of electricity consumption, the models show significant adaptation throughout the day. Shorter autoregressive orders ($p=1,2,4$) are selected at various hours, indicating that while weekly patterns are important, shorter-term dynamics become more relevant at specific times. For the moving average component, both $q=1$ and $q=2$ are frequently included reflecting flexibility in capturing short-term forecast errors.

The results highlighted in Table \ref{LBMs} are confirmed by the findings in Figure \ref{IIs}, where we show the normalized inclusion importance introduced in Equation (\ref{eq:normalized_II}) for key predictor variables at different hours. The contribution of each variable to explaining load variation is clearly time-dependent. Moreover, the close agreement between inclusion importance at the confidence set level ($\widetilde{II}_s$) and at the LBM level ($II$) indicates that the main findings on intra-day hourly loads, calendar effects, and temperature are driven by features common to the whole confidence set, rather than by the choice of any particular parsimonious specification.

\begin{table}
\centering
\begin{adjustbox}{width=\textwidth,keepaspectratio}
\begin{tabularx}{\textwidth}{ll*{9}{>{\centering\arraybackslash}X}}
\toprule
\multicolumn{2}{c}{\textbf{Model term}} & \multicolumn{9}{c}{\textbf{Hour}} \\
\cmidrule(lr){3-11}
& & 8 & 9 & 10 & 11 & 12 & 13 & 14 & 15 & 16 \\
\midrule
\multicolumn{2}{c}{} & \multicolumn{9}{c}{$\alpha = 0.01$} \\
\midrule
& \texttt{Lag$_{1}$} & \cellcolor{gray!100} & \cellcolor{gray!100} & \cellcolor{gray!100} & \cellcolor{gray!100} & \cellcolor{gray!100} & \cellcolor{gray!100} & \cellcolor{gray!100} & \cellcolor{gray!100} & \cellcolor{gray!100} \\
& \texttt{Lag$_{2}$} & \cellcolor{gray!100} & \cellcolor{gray!100} & \cellcolor{gray!100}& \cellcolor{gray!100}&\cellcolor{gray!33} &\cellcolor{gray!100} & \cellcolor{gray!100} & \cellcolor{gray!100} & \cellcolor{gray!100} \\
& \texttt{Lag$_{3}$}& & \cellcolor{gray!100} & & & \cellcolor{gray!100} & \cellcolor{gray!100} & \cellcolor{gray!100} & & \cellcolor{gray!33} \\
& \texttt{Week/Hol.} & \cellcolor{gray!100} & \cellcolor{gray!100} & & \cellcolor{gray!100} & \cellcolor{gray!100} & \cellcolor{gray!100}& \cellcolor{gray!100} & \cellcolor{gray!100} & \\
& \texttt{Temperature} & & \cellcolor{gray!100}&\cellcolor{gray!100} & \cellcolor{gray!100} & \cellcolor{gray!100} & & \cellcolor{gray!20} & \cellcolor{gray!33} & \cellcolor{gray!33} \\
& \texttt{Temperature$^2$} & & \cellcolor{gray!100} & \cellcolor{gray!100} & \cellcolor{gray!100} & \cellcolor{gray!100} & \cellcolor{gray!100} & \cellcolor{gray!20} & \cellcolor{gray!33} & \cellcolor{gray!66} \\
& \texttt{Solar} & \cellcolor{gray!100} & \cellcolor{gray!100} & \cellcolor{gray!100} &\cellcolor{gray!100} &\cellcolor{gray!100} &\cellcolor{gray!33}& \cellcolor{gray!80} & \cellcolor{gray!33} & \cellcolor{gray!66} \\
& \texttt{Wind} & & & & &&\cellcolor{gray!66}&\cellcolor{gray!80}&&\cellcolor{gray!33}\\
& \texttt{Summer}&&&&\cellcolor{gray!100}&\cellcolor{gray!100}&\cellcolor{gray!33}&\cellcolor{gray!33}&&\cellcolor{gray!33}\\
& \texttt{Winter} && \cellcolor{gray!50}&& \cellcolor{gray!100} &\cellcolor{gray!33}&\cellcolor{gray!100}&\cellcolor{gray!20}&\\
& \texttt{Lockdown} && \cellcolor{gray!100}&&&& &&\cellcolor{gray!33}&\\
& \texttt{y$_{t-1}$}  &\cellcolor{gray!100}&\cellcolor{gray!100}&\cellcolor{gray!100}&\cellcolor{gray!100}&\cellcolor{gray!100}&\cellcolor{gray!100}&\cellcolor{gray!100}&\cellcolor{gray!100}&\cellcolor{gray!100} \\
& \texttt{y$_{t-2}$}&\cellcolor{gray!100}&\cellcolor{gray!100}&\cellcolor{gray!100}&\cellcolor{gray!100}&\cellcolor{gray!33}&\cellcolor{gray!100}&\cellcolor{gray!100}&\cellcolor{gray!33}&\cellcolor{gray!100}\\
& \texttt{y$_{t-3}$}&\cellcolor{gray!100}&\cellcolor{gray!100}&\cellcolor{gray!100}&\cellcolor{gray!100}&&\cellcolor{gray!100}&\cellcolor{gray!100}&&\cellcolor{gray!100}\\
& \texttt{y$_{t-4}$} &\cellcolor{gray!100}&\cellcolor{gray!100}&\cellcolor{gray!100}&\cellcolor{gray!100}&&\cellcolor{gray!100}&\cellcolor{gray!100}&&\cellcolor{gray!100}\\
& \texttt{y$_{t-5}$}&\cellcolor{gray!100}&\cellcolor{gray!100}&\cellcolor{gray!100}&\cellcolor{gray!100}&&\cellcolor{gray!100}&\cellcolor{gray!40}&&\cellcolor{gray!100}\\
& \texttt{y$_{t-6}$} &\cellcolor{gray!100}&\cellcolor{gray!100}&\cellcolor{gray!100}&\cellcolor{gray!100}&&\cellcolor{gray!100}&\cellcolor{gray!40}&&\cellcolor{gray!100}\\
& \texttt{y$_{t-7}$}&\cellcolor{gray!100}&\cellcolor{gray!100}&\cellcolor{gray!100}&\cellcolor{gray!100}&&\cellcolor{gray!100}&\cellcolor{gray!40}&&\cellcolor{gray!100}\\
& \texttt{$\bm{\varepsilon}_{t-1}$}&\cellcolor{gray!100}&\cellcolor{gray!100}&\cellcolor{gray!100}&\cellcolor{gray!100}&\cellcolor{gray!100}&\cellcolor{gray!100}&\cellcolor{gray!100}&\cellcolor{gray!100}&\cellcolor{gray!100}\\
&\texttt{$\bm{\varepsilon}_{t-2}$}&\cellcolor{gray!100}&\cellcolor{gray!50}&&&\cellcolor{gray!66}&\cellcolor{gray!33}&\cellcolor{gray!60}&\cellcolor{gray!66}\\
\midrule
\multicolumn{2}{c}{} & \multicolumn{9}{c}{$\alpha = 0.05$} \\
\midrule
& \texttt{Lag$_{1}$} & \cellcolor{gray!100} & \cellcolor{gray!100} & \cellcolor{gray!100} & \cellcolor{gray!100} & \cellcolor{gray!100} & \cellcolor{gray!100} & \cellcolor{gray!100} & \cellcolor{gray!100} & \cellcolor{gray!100} \\
& \texttt{Lag$_{2}$} & \cellcolor{gray!100} & \cellcolor{gray!100} & \cellcolor{gray!100}& \cellcolor{gray!100}&\cellcolor{gray!66} &\cellcolor{gray!100} & \cellcolor{gray!100} & \cellcolor{gray!100} & \cellcolor{gray!100} \\
& \texttt{Lag$_{3}$} & & \cellcolor{gray!100} & & & \cellcolor{gray!100} & \cellcolor{gray!100} & \cellcolor{gray!100} & & \cellcolor{gray!33} \\
& \texttt{Week/Hol.} & \cellcolor{gray!100} & \cellcolor{gray!100} & \cellcolor{gray!100} & \cellcolor{gray!100} & \cellcolor{gray!100} & \cellcolor{gray!100}& \cellcolor{gray!100} & \cellcolor{gray!100} & \\
& \texttt{Temperature} & \cellcolor{gray!33}& \cellcolor{gray!100}& \cellcolor{gray!100} & \cellcolor{gray!100} & \cellcolor{gray!100} & & \cellcolor{gray!51} & \cellcolor{gray!50} & \cellcolor{gray!66} \\
& \texttt{Temperature$^2$} & \cellcolor{gray!33} & \cellcolor{gray!100} & \cellcolor{gray!100} & \cellcolor{gray!100} & \cellcolor{gray!100} & \cellcolor{gray!100} & \cellcolor{gray!51} & \cellcolor{gray!50} & \cellcolor{gray!100} \\
& \texttt{Solar} & \cellcolor{gray!100} & \cellcolor{gray!100} & \cellcolor{gray!100} &\cellcolor{gray!100} &\cellcolor{gray!100} &&\cellcolor{gray!100}& \cellcolor{gray!25} & \cellcolor{gray!66} \\
& \texttt{Wind} & & & & &&\cellcolor{gray!50}&\cellcolor{gray!97}&&\\
& \texttt{Summer}&&\cellcolor{gray!50}&&\cellcolor{gray!100}&\cellcolor{gray!100}&\cellcolor{gray!50}&\cellcolor{gray!55}&\cellcolor{gray!25}&\cellcolor{gray!66}\\
& \texttt{Winter} && \cellcolor{gray!100}&& \cellcolor{gray!100} &\cellcolor{gray!66}&\cellcolor{gray!100}&\cellcolor{gray!47}&\cellcolor{gray!25}&\\
& \texttt{Lockdown} &\cellcolor{gray!33}& \cellcolor{gray!100}&&&& &\cellcolor{gray!45}&\cellcolor{gray!75}&\\
& \texttt{y$_{t-1}$} &\cellcolor{gray!100}&\cellcolor{gray!100}&\cellcolor{gray!100}&\cellcolor{gray!100}&\cellcolor{gray!100}&\cellcolor{gray!100}&\cellcolor{gray!100}&\cellcolor{gray!100}&\cellcolor{gray!100}\\
& \texttt{y$_{t-2}$} &\cellcolor{gray!100}&\cellcolor{gray!100}&\cellcolor{gray!100}&\cellcolor{gray!100}&\cellcolor{gray!100}&\cellcolor{gray!100}&\cellcolor{gray!100}&\cellcolor{gray!25}&\cellcolor{gray!100}\\
& \texttt{y$_{t-3}$}  &\cellcolor{gray!100}&\cellcolor{gray!100}&\cellcolor{gray!100}&\cellcolor{gray!100}&\cellcolor{gray!66}&\cellcolor{gray!100}&\cellcolor{gray!100}&&\cellcolor{gray!100}\\
& \texttt{y$_{t-4}$} &\cellcolor{gray!100}&\cellcolor{gray!100}&\cellcolor{gray!100}&\cellcolor{gray!100}&\cellcolor{gray!66}&\cellcolor{gray!100}&\cellcolor{gray!100}&&\cellcolor{gray!100}\\
& \texttt{y$_{t-5}$} &\cellcolor{gray!100}&\cellcolor{gray!100}&\cellcolor{gray!100}&\cellcolor{gray!100}&\cellcolor{gray!66}&\cellcolor{gray!100}&\cellcolor{gray!60}&&\cellcolor{gray!100}\\
& \texttt{y$_{t-6}$}  &\cellcolor{gray!100}&\cellcolor{gray!100}&\cellcolor{gray!100}&\cellcolor{gray!100}&\cellcolor{gray!66}&\cellcolor{gray!100}&\cellcolor{gray!60}&&\cellcolor{gray!100}\\
& \texttt{y$_{t-7}$}  &\cellcolor{gray!100}&\cellcolor{gray!100}&\cellcolor{gray!100}&\cellcolor{gray!100}&&\cellcolor{gray!100}&\cellcolor{gray!60}&&\cellcolor{gray!100}\\
&\texttt{$\bm{\varepsilon}_{t-1}$}&\cellcolor{gray!100}&\cellcolor{gray!100}&\cellcolor{gray!100}&\cellcolor{gray!100}&\cellcolor{gray!100}&\cellcolor{gray!100}&\cellcolor{gray!100}&\cellcolor{gray!100}&\cellcolor{gray!100}\\
& \texttt{$\bm{\varepsilon}_{t-2}$}&\cellcolor{gray!100}&\cellcolor{gray!50}&&&&\cellcolor{gray!100}&\cellcolor{gray!40}&\cellcolor{gray!75}\\

\bottomrule
\end{tabularx}
\end{adjustbox}
\caption{Inclusion importance ($II$) of different model terms at 99\% and 95\%  confidence level for working hours. Darker cells indicate larger inclusion importance. }\label{LBMs}
\end{table}

\subsection{Evaluating ARMAX models selected by common criteria}

Next, we use the MSCS to assess whether models selected by common information criteria are trustworthy. Table~\ref{tab:sel} compares predictor inclusion patterns between our method and the single best model chosen by three conventional criteria: the Akaike Information Criterion (AIC), the Bayesian Information Criterion (BIC), and the model minimizing out-of-sample Root Mean Squared Error (RMSE). We examine hours 8, 12, and 16 to capture morning, midday, and evening load dynamics.
The analysis reveals substantial temporal variation in whether criterion-selected models fall within the MSCS. At hours 8 and 12, all three criteria—AIC, BIC, and RMSE—select models that are included within the MSCS at the 99\% confidence level, indicating that these models are statistically plausible during morning and midday periods. However, at hour 16, both BIC and RMSE identify models outside the MSCS bounds, while only AIC selects a model within our confidence set.
This result reflects the fundamental trade-offs embedded in these selection criteria. AIC's weaker penalty for model complexity tends to favor models with more parameters, which increases the likelihood that AIC-selected models fall within the MSCS when the true data-generating process exhibits substantial structure. Conversely, BIC's stronger parsimony penalty favors more parsimonious specifications, which may fall outside the MSCS when the underlying load dynamics require additional complexity to achieve statistical plausibility. The divergence at hour 16 suggests that evening load patterns exhibit greater complexity than earlier periods, requiring a richer model structure that BIC's penalty excludes but that the MSCS deems statistically necessary. On the other hand, RMSE's focus on minimizing prediction error, rather than balancing fit and complexity through an information-theoretic penalty, can similarly lead to underfitted models that lack the statistical support required for MSCS inclusion. These findings adds to the value of the MSCS in providing a rigorous benchmark for evaluating whether conventional criteria strike an appropriate balance between parsimony and model adequacy.
Furthermore, examining individual predictors reveal nuanced temporal patterns. Among intraday lags, the one-hour lag (\texttt{Lag\_1}) appears consistently across all methods and hours, while the two-hour and three-hour lags (\texttt{Lag\_2}, \texttt{Lag\_3}) display more variable selection patterns, with inclusion rates changing across different hours and selection criteria.

\texttt{Temperature} and its squared term exhibit time-varying importance: largely excluded at hour 8 but frequently selected at hours 12 and 16, consistent with the thermal sensitivity of energy consumption during warmer parts of the day. \texttt{Solar}  shows strong inclusion at hour 8 but diminishing importance at hours 12 and 16, reflecting its natural availability pattern.  \texttt{Week./Hol.}  indicator maintains high inclusion during morning and midday but loses significance at hour 16, possibly due to convergence between weekday and weekend load patterns during evening hours. Seasonal indicators (\texttt{Summer}, \texttt{Winter}) show moderate inclusion primarily at hour 12, while the \texttt{Lockdown} indicator appears only sporadically, selected solely by RMSE models at hours 12 and 16.
The ARMA structure parameters provide further insight into temporal modeling requirements. The weekly lag  $y_{t-7}$ is selected across most methods and hours, confirming strong weekly seasonality in energy load. The MA structure varies by hour: $\varepsilon_{t-2}$ is required at hour 8, whereas $\varepsilon_{t-1}$ alone suffices at later hours.

\begin{table}[h!]
    \centering
    \begin{adjustbox}{width=\textwidth, keepaspectratio} 
    \label{sel}
    \begin{tabular}{l ccccc | ccccc | ccccc}
        \toprule
        & \multicolumn{5}{c|}{\textbf{Hour 8}} & \multicolumn{5}{c|}{\textbf{Hour 12}} & \multicolumn{5}{c}{\textbf{Hour 16}} \\
        \cmidrule(lr){2-6} \cmidrule(lr){7-11} \cmidrule(lr){12-16}
        \textbf{Model term} & $\widetilde{II} _s$ & ${II}$ & AIC & BIC & RMSE & $\widetilde{II}_s$ & ${II}$ & AIC & BIC & RMSE & $\widetilde{II}_s$ & ${II}$ & AIC & BIC & RMSE \\
        \midrule
        \texttt{Lag$_{1}$} & 1 & 1 & 1 & 1 & 1 & 1 & 1 & 1 & 1 & 1 & 1 & 1 & 1 & 1 & 1 \\
         \texttt{Lag$_{2}$} & 1& 1 & 1 & 1 & 1 &0.4 & 0.3 & 1 & 0 &1  & 1 & 1  & 1 & 1 & 0   \\
         \texttt{Lag$_{3}$} & 0&  0&  0&  0& 0 &1 & 1 & 1 & 1 & 1 & 0.1 & 0.3 & 1 & 0 & 1   \\
        \texttt{Week.\&Hol.}&1 & 1 & 1 & 1 & 1 &1 & 1 & 1 & 1 & 1 & 0 & 0 &0 & 0 &  1  \\
        \texttt{Temperature} & 0&  0&  0&  0& 0 &1 & 1 & 1 & 1 & 1 & 0.6 & 0.3 &  1& 0 & 1   \\
        \texttt{Temperature$^2$} &0 &0 &1 & 1 & 1 &1 & 1 & 1 & 1 & 1 & 1 & 0.7 & 1 & 0 &  0  \\
       \texttt{Solar} &1& 1 &1  &1  &1  & 1 & 1 & 1 & 1 & 1 & 0.4 & 0.7 & 1 & 0 &  0  \\
       \texttt{Wind} & 0& 0 & 0 & 0 & 0 &0 & 0 & 1 & 0 & 0 & 0 & 0.3 & 0 & 0 &  1  \\
        \texttt{Summer} &0 &0  &0  &0  & 0 &1 & 1 & 1 & 1 & 1 & 0.3 & 0.3 & 1 & 0 & 0   \\
        \texttt{Winter} & 0&0  &0  & 0 & 0 &0.2 & 0.3 & 1 & 0 & 1 & 0 & 0 & 0 & 0 &  0  \\
        \texttt{Lockdown} &0&  0&  0&  0& 0 & 0 & 0 & 0 & 0 & 1 & 0 & 0 & 0 & 0&  1  \\
        \midrule
        \texttt{y$_{t-1}$} & 1 & 1 & 1 & 1 & 1& 1 & 1 & 1& 1 & 1 & 1 & 1 & 1 & 1 &  1  \\
        \texttt{y$_{t-2}$}&1&  1&  1&  1&  1& 0 & 0.3 & 1 & 1 & 1 & 1 & 1 & 1 & 1 &  1 \\
        \texttt{y$_{t-3}$}  &1  &1  &1  &1  &1& 0 & 0 & 1 & 0 & 1 & 1 & 1 & 1 & 1 &  1 \\
        \texttt{y$_{t-4}$} & 1& 1 &1  &1&1  &0 & 0 & 1 & 0 & 1 & 1 & 1 & 1 & 1 &  1  \\
        \texttt{y$_{t-5}$} & 1 & 1 & 1 & 1 & 1 &0 & 0 & 1 & 0 & 1 & 1 & 1 & 1 & 1 &  1 \\
        \texttt{y$_{t-6}$} & 1& 1 & 1 & 1 & 1 &0 & 0 & 1 & 0 & 1 & 1 & 1 & 1 & 1 &  1  \\
        \texttt{y$_{t-7}$} &1& 1 & 1 & 1 & 1 & 0 & 0 & 1 & 0 & 0 & 1 & 1 & 1 & 1 &  1  \\
        \texttt{$\bm{\varepsilon}_{t-1}$}
        & 1& 1 & 1 & 1 & 1 &1 & 1 & 1 & 1 & 1 & 1 & 1 & 1 & 1 & 1   \\
        \texttt{$\bm{\varepsilon}_{t-2}$} &1 & 1 & 1 & 1 & 1 &0 & 0.7 & 0 & 0 & 0 & 0 & 0 & 0 & 0 &  0  \\
        \midrule
        Included in MSCS & - & - & Yes & Yes & Yes & - & - & Yes & Yes & Yes & - & - & Yes & No & No \\
        \bottomrule
    \end{tabular}
    \end{adjustbox} 
    \caption{Normalized inclusion importance based on the Model Selection Confidence Set (MSCS) ($\widetilde{II}_s$) and inclusion importance ($II$) derived from LBMs  for exogenous predictors and orders, computed at the 99\% confidence level for time series observed at hour 8,  12, and 16. Additional columns indicate the predictors included in models that minimize AIC, BIC, and root mean squared error (RMSE) criteria ($1=\text{selected}$, $0=\text{excluded}$).}
    \label{tab:sel}
\end{table}

\subsection{Predictive accuracy of MSCS}

Given the ability of the MSCS to identify models statistically equivalent to the true data-generating process, we study whether our methodology also ensures the selection of well-performing forecasting models. We evaluate this by comparing the predictive performance of models inside the MSCS with the ones left outside. Accuracy is measured using out-of-sample point forecast metrics for hourly load, specifically the root-mean-square error and the mean absolute error:

\begin{equation}
\text{RMSE}_h = \sqrt{\frac{1}{T - R} \sum_{t=R}^{T-1} \left(\hat{y}_{h,t+1|t} - y_{h,t+1}\right)^2},
\qquad
\text{MAE}_h = \frac{1}{T - R} \sum_{t=R}^{T-1} \left|\hat{y}_{h,t+1|t} - y_{h,t+1}\right|,
\end{equation}
where $T$ is the number of observations, $R$ is the length of the rolling window and $\hat{y}_{h,t+1|t}$
are the individual hourly load forecasts.
Specifically, results are based on a one-step ahead rolling forecasting process with a window size
of three years. The initial estimation sample goes from 1 January 2019 to
31 December 2021 and then the forecasting evaluation period starts on 1 January 2023
and ends on 29 September 2023; for a total of 637 out-of-sample point forecasts.

\begin{figure}
    \centering
    \includegraphics[width=1.0\linewidth]{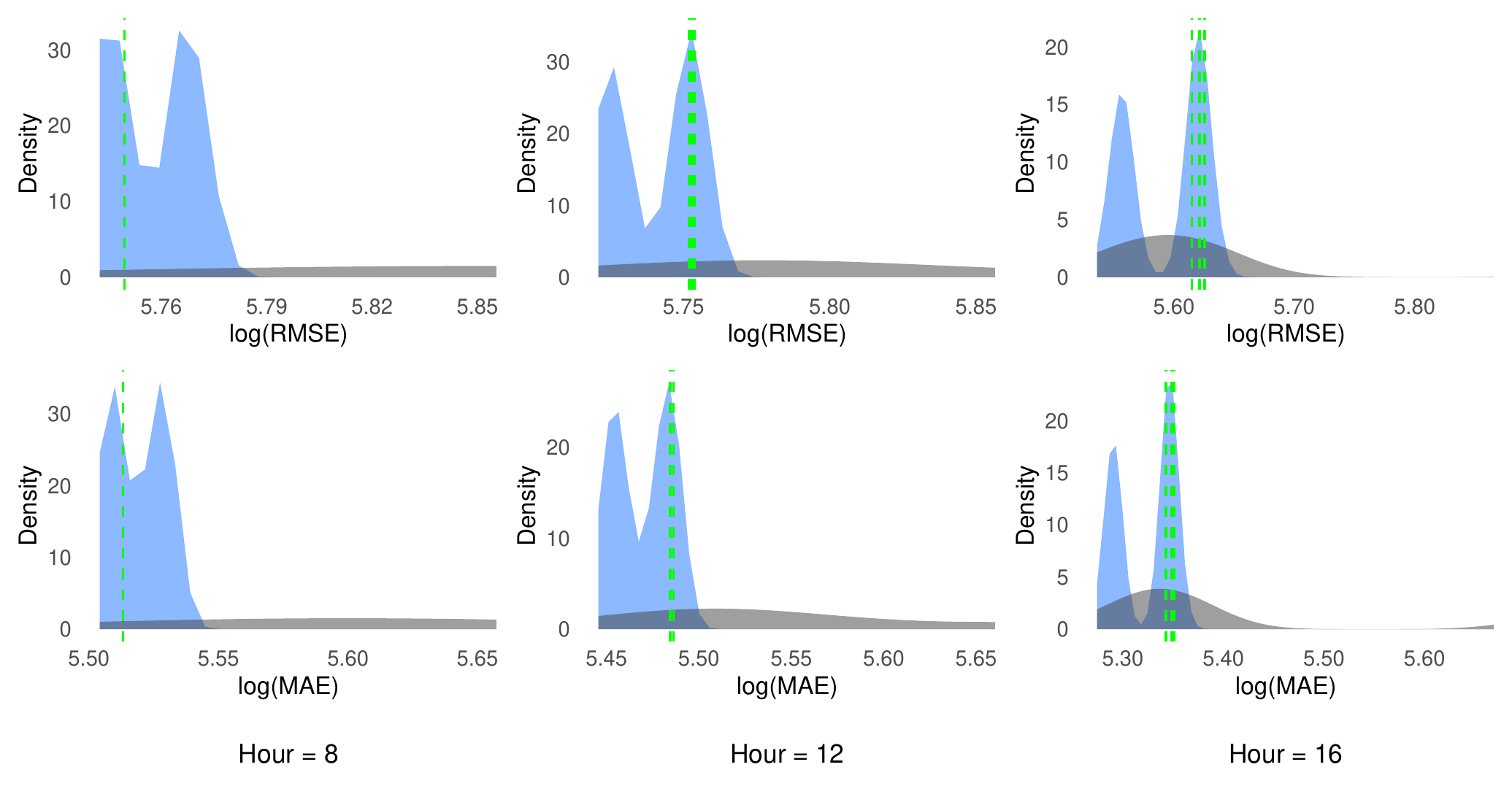}
    \caption{Distributions of logarithmic RMSE and MAE for models in the confidence set (blue curve) versus models out the MSCS (gray curve) computed at the 99\% confidence level for time series observed at hour 8,  12, and 16. Vertical green dashed lines represent the performance of LBMs.}
    \label{distributions}
\end{figure}

Figure \ref{distributions} compares the out-of-sample forecast error distributions (RMSE and MAE) at hours 8, 12, and 16 for models that are included in the MSCS and for those that are excluded. Both metrics convey a consistent message: excluded models exhibit a wide dispersion of forecast errors, including many specifications with substantially poorer performance, whereas models in the MSCS are tightly concentrated in the low-error region. This clear separation indicates that the MSCS isolates a set of model specifications that deliver competitive short-term forecasts of electricity load, and thus provides a useful diagnostic for model choice in forecasting applications. The contrast in the distributions mirrors the MSCS selection mechanism, which is designed to retain models whose losses are statistically indistinguishable from the best-performing specification while discarding models with significantly worse performance.

The results for the LBMs further illustrate the trade-off between parsimony and predictive accuracy. At hour 8, the LBMs rank among the most accurate models, while at hours 12 and 16 they remain highly competitive but lie closer to the right tail of the MSCS distribution. This suggests that, although not always close to the empirical optimum, LBMs offer a good compromise between model parsimony and forecasting performance. 

Overall, these findings provide empirical support that the models retained in the confidence set also perform well out of sample, reinforcing the practical usefulness of the MSCS framework for prediction tasks.

\section{Conclusion}\label{sec:conclusion}
Quantifying uncertainty in model selection is becoming a central topic in time series analysis and forecasting. In this paper, we introduced and applied the Model Selection Confidence Set (MSCS) methodology for ARMAX models. The key idea is to move beyond the traditional single-model paradigm and provide, instead, a statistically valid set of admissible specifications. In doing so, the MSCS makes selection variability explicit and provides a principled way to assess structural uncertainty in dynamic environments where misspecification risks are high.

Monte Carlo experiments illustrated the finite-sample behavior of the approach under different ARMAX structures. As sample size increases, the MSCS attains appropriate coverage probabilities and the size of the corresponding LBM sets contracts. Moreover, simpler AR structures tend to be identified more quickly than models that include additional moving-average and exogenous components.

To summarize stability across admissible models, we proposed inclusion importance statistics that quantify how frequently terms appear within the LBMs and across the entire confidence set. In the application to Italian electricity load data we find that selection uncertainty varies systematically over the day: intraday lags, temperature, calendar effects, and solar generation emerge consistently as key drivers, while the overall degree of structural uncertainty is time-varying with morning hours yielding smaller, more stable confidence sets, whereas afternoons admitting a broader range of plausible specifications. 
This highlights the practical value of acknowledging selection uncertainty when modeling and forecasting electricity load.

Looking ahead, applications to other commodity markets or energy sectors will likely involve much larger predictor pools and richer lag structures, making exhaustive enumeration of the model space infeasible. A natural direction is therefore to combine the MSCS with a conservative screening step that reduces the candidate pool to a manageable subset, treat that reduced specification as the “full model,” and then construct the confidence set over nested models within the restricted space. In addition, because many applications operate at sample sizes where asymptotic approximations may be inaccurate, future work should investigate bootstrap likelihood ratios or other resampling-based calibrations to improve finite-sample coverage and sharpen uncertainty quantification.

\bibliographystyle{apalike} 
\bibliography{references}

 \end{document}